\documentclass[review]{elsarticle}
\usepackage{bm}
\usepackage{graphicx}
\usepackage{amssymb}
\usepackage{amsmath}

\usepackage{lineno,hyperref}
\modulolinenumbers[5]

\journal{Physics Letters B}

\begin{document}

\begin{frontmatter}

\title{
Exact scattering waves off nonlocal potentials 
under Coulomb interaction
within Schr\"odinger's integro--differential equation
}

\author{H. F. Arellano \fnref{mymainaddress,mysecondaryaddress}}
\author{G. Blanchon \fnref{mysecondaryaddress}}

\address[mymainaddress]{Department of Physics - FCFM, University of Chile\\
	Av. Blanco Encalada 2008, Santiago, Chile}
\address[mysecondaryaddress]{CEA,DAM,DIF, F-91297 Arpajon, France}

\begin{abstract}
An exact solution for the scattering wavefunction from a nonlocal 
potential in the presence of Coulomb interaction is presented. 
The approach is based on the construction of a Coulomb Green's function 
in coordinate space whose associated kernel involves any
nonlocal optical potential superposed to the 
Coulomb--screened interaction. 
The scattering wavefunction, exact solution of the 
integro--differential Schr\"odinger's equation,
poses no restrictions on the type of nonlocality of the 
interaction nor on the beam energy.
\end{abstract}

\begin{keyword}
        Nonlocal potential\sep 
	scattering theory\sep 
	optical model potential \sep 
	integro-differential equation \sep
        Coulomb interaction
\end{keyword}

\end{frontmatter}


\section{Introduction}
Beyond its intrinsic merit, the value of counting on an exact solution to any given 
problem is that it provides with accuracy benchmarks for alternative approaches.
In the particular case of the interaction of a single nucleon with a nucleus it is 
well established that the coupling is nonlocal, feature that arises from the fermionic 
nature of all interacting nucleons. In the presence of a nonlocal potential Schr\"odinger's 
equation for scattering waves becomes integro-differential. 
Explicit treatments of nonlocalities in Schr\"odinger's equation is an issue 
that has captured increasing interest from the stand point of {\it ab-initio} 
theories and models, specially aiming to global approaches for structure and 
reactions \cite{Rotureau2017,Idini2018,Dickhoff2017}. Therefore, robust methods 
able to provide solutions to the wave equation for any kind of kernel become imperative 
to accurately treat and assess model-independent nonlocalities of nuclear interactions. 
To this date, however, it can safely be stated that the only established kernel--independent 
approach that solves exactly Schr\"odinger's equation for the wavefunction --in the presence 
of Coulomb interaction--is the one reported in Refs. \cite{Raynal1998,Gibbs2006}. The method is 
based on finite difference techniques, reducing the problem to a matrix equation for 
the wavefunction. In this work we present an alternative solution to the integro--differential
equation, resulting in a non--singular integral equation readily invertible. The key 
feature in this case is the construction of a Green's function capable of accounting 
exactly for the underlying long--range Coulomb interaction.

In the context of nucleon--nucleus scattering, physical quantities of major interest 
are the scattering amplitudes and wavefunctions. The latter being useful, for example, 
in distorted wave Born approximation applications. When expressed in coordinate space 
the equation for the wavefunction becomes integro--differential. Early solutions to this 
problem were proposed by Perey and Buck \cite{Perey1962}, transforming the non-local 
potential by a local--equivalent. A shortcoming of this approach is that the calculated 
outgoing wavefunction differs from the exact one, distortion which is known as Perey effect 
and characterized by the Perey correction factor \cite{Titus2014}.

Other solutions to Schr\"odinger's integro--differential equation follow iterative 
procedures \cite{Titus2014,Titus2016}. In these schemes Schr\"odinger's differential equation 
is integrated with a non--homogeneous term consisting of the projection of the nonlocal potential 
onto an intermediate solution, 
$U_{nl}|\chi_i\rangle$. 
These procedures 
begin with a given seed to generate the starting solution $|\chi_0\rangle$, with subsequent 
iterations until convergence is reached. These iterative methods may require prior knowledge 
of the solution in order to make convergence more efficient, though there is no guarantee 
to converge to the correct solution. In the case of Ref. \cite{Upadhyay2018}, a mean--value 
technique is applied to approximate $U_{nl}|\chi_i\rangle$, reducing the problem to a second--order 
homogeneous differential equation. Quite recently another approach has been proposed to deal with 
nonlocal potentials \cite{Upadhyay2018b}, where a Taylor approximation for the radial wave function 
is applied. This strategy is based on the assumption that nonlocality is dominant around the diagonal 
in coordinate space, a non universal feature as reported in Ref.~\cite{Arellano2018} for microscopic 
potentials based on off--shell $g$ matrices.

Another method to calculate waves off nonlocal potentials in the presence of long--range Coulomb 
interaction is that of Refs.~\cite{Kim1990,Kim1992}, where Lanczos technique is used to solve integral 
equations derived from the nonlocal Schr\"odinger equation. More recently, in Refs.~\cite{Hagen2012,Michel2011} 
a numerical treatment to this problem has been proposed with the use of Berggren basis, where an 
off--diagonal approximation is used to control the Coulomb singularity along the diagonal in momentum space. 
Applications of this approach have been restricted to low energies and intermediate mass targets.

Solutions to the scattering problem in momentum space have also been investigated 
\cite{Picklesimer1984,Arellano1990,Elster1990,Crespo1990,Chinn1991,Lu1994,Upadhyay2014}. 
See Ref.~\cite{Ray1992} for a review on the subject. While an advantage of momentum--space approaches is that 
nonlocalities are naturally accounted for, one of its drawbacks is that no method
is available to extract the associated scattering waves. In the absence of Coulomb interaction the calculation of 
scattering amplitudes is rather straightforward, reducing the problem to a Lippmann--Schwinger integral equation 
for the scattering matrix. However, in the presence of Coulomb potential the approach
cannot be applied right away due to the $\sim$$1/q^{2}$ singularity of the interaction.
An exact solution addressing this singularity has been proposed by Vincent and Phatak by means of a 
cut--off technique to the Coulomb long--range tail \cite{Vincent1974}. This approach has been applied 
to proton--nucleus scattering at intermediate energies \cite{Arellano1990}, where its accuracy is 
significantly improved after a detailed multipole treatment of the charge form factor convoluted with 
a sharp cut--off potential \cite{Eisenstein1982}. 

In this article we present exact solutions for scattering waves off any finite--range nonlocal potential 
in coordinate space, where the Coulomb interaction is included without approximation. The approach, briefly 
sketched in an appendix of Ref.~\cite{Arellano2007b} in the context of quasielastic $(p,n)$ charge--exchange 
reactions, is not restricted on energy of the projectile, charge of the colliding particles nor nature of the 
nonlocality. 

This paper is organized as follows. In Sec.~\ref{coulomb} we lay out the framework and present a formal solution 
to the scattering problem with nonlocal potentials in the presence of Coulomb interactions. We provide a demonstration 
of the solution and illustrate its consistency with a numerical example.
In Sec.~\ref{conclu} we present the main conclusions of the work.

\section{Integral equation for scattering waves}
\label{coulomb}
Let us consider the collision of a proton with a nucleus of charge $Ze$. 
The interaction $U$ between them is given by the sum of a pure 
hadronic contribution ($U_H$) and the Coulomb interaction ($U_C$) due 
to the charge distribution of the nucleus, $U=U_H+U_C$. 
The hadronic part is regarded in general as a nonlocal operator so that
the total potential can be cast as the sum of a point--Coulomb and 
short--range terms,
\begin{equation}
  \label{us}
U({\bf r}',{\bf r}) = 
U^{[s]}({\bf r}',{\bf r}) + \frac{\beta}{r}\delta({\bf r}'-{\bf r})\;,
\end{equation}
with $\beta=Ze^2$.
Here $U^{[s]}$ defines the finite--range part of the interaction
where the point--Coulomb interaction has been subtracted,
namely $U^{[s]}=U_H+U_C-\beta\delta({\bf r}'-{\bf r})/r$.

With the above construction in mind we examine
Schr\"odinger's equation for scattering waves,
which in coordinate representation reads
\begin{equation}
\label{schr}
-\nabla^2 \psi_{\bf k} ({\bf r}) + 
\frac{2m}{\hbar^2} 
\int d{\bf r'} U({\bf r},{\bf r'})\psi_{\bf k}({\bf r'})=
k^2 \psi_{\bf k}({\bf r'}) \; ,
\end{equation}
with $m$ the nucleon--nucleus reduced mass and
$k$ the asymptotic relative momentum. Spin and isospin variables 
are omitted for notation simplicity.
Considering a spin-$0$ closed--shell target interacting with a
spin-$\textstyle{\frac12}$ nucleon, 
the following partial wave expansion for the 
scattering wavefunction becomes suitable,
\begin{equation}
\label{expansion}
\psi_{\bf k} ({\bf r}) =
  \sqrt{\frac{2}{\pi}}\;
  \sum_{jlm} i^l 
{\cal Y}_{jl1/2}^{m}({\bf \hat r})
e^{i\sigma_l}
\frac{u_{jl}(r)}{r} 
{\cal Y}_{jl1/2}^{m\dagger}({\bf \hat k}) \; .
\end{equation}
In this expansion ${\cal Y}_{jl1/2}^{m}$ denotes spherical vectors and $\sigma_l$ the Coulomb 
phase--shift for partial wave $l$. Here $u_{jl}(r)$ is the radial wavefunction. In the limit where 
the finite--range interaction $U^{[s]}$ is set to zero, 
the unperturbed wavefunction becomes a
free Coulomb wave due to a pointlike source, 
$\psi_{\bf k} ({\bf r})~\to~\phi_{c}({\bm r})$, 
where
\begin{equation}
\label{freecoulomb}
\phi_{c}({\bf r})
=
  \sqrt{\frac{2}{\pi}}\;
  \sum_{jlm} i^l
{\cal Y}_{jl1/2}^{m}({\bf \hat r}) e^{i\sigma_l}
F_{l}(kr) {\cal Y}_{jl1/2}^{m\dagger}({\bf \hat k}) \; ,
\end{equation}
with $F_l$ the regular Coulomb function.
In the absence of Coulomb interaction ($\beta=0$), 
this expression leads to normalized plane waves  $\phi_{\bf k} ({\bf r})$,
\begin{equation}
\phi_{\bf k} ({\bf r}) = 
\frac{\bm 1_\sigma}{(2\pi)^{3/2}}e^{i{\bf k}\cdot{\bf r}}\;,
\end{equation}
with ${\bm 1_\sigma}$ the identity in spin--$\textstyle{\frac12}$--space.

By replacing $\psi_{\bf k} ({\bf r})$ from Eq.~\eqref{expansion}
into Eq.~\eqref{schr}, following standard procedures we obtain \cite{Joachain1975}
\begin{align}
\left [
\frac{1}{r}
\left ( 
\frac{d^2}{dr^2}
\right ) r - \frac{l(l+1)}{r^2} + k^2
\right ] \frac{u_{jl}(r)}{r} = 
\frac{2m}{\hbar^2}
\int_0^\infty r'\,dr'
U_{jl}(r,r')
u_{jl}(r') \; ,
\end{align}
where the multipoles $U_{jl}$ of the interaction are obtained from
\begin{equation}
U_{jl}(r',r) = \iint d\hat{\bf r}\,d\hat{\bf r}' 
{\cal Y}_{jl1/2}^{m \dagger}({\bf \hat r}') 
U({\bf r'},{\bf r})
{\cal Y}_{jl1/2}^{m}({\bf \hat r}) \;.
\end{equation}
Making explicit the separation of the interaction
into a pointlike source and finite--range remaining
\begin{equation}
U_{jl}(r',r) \equiv U_{jl}^{[s]}(r',r) + \frac{\beta}{r^3}\delta(r'-r) \; ,
\end{equation}
we obtain
\begin{align}
\label{schr2}
{\cal D}_{c} u_{jl}(r) 
& \equiv
\left [
\frac{d^2}{dr^2} - \frac{l(l+1)}{r^2} -\frac{2k\eta}{r}+ k^2
\right ] u_{jl}(r) 
\nonumber \\
& =
\frac{2m}{\hbar^2} 
\int dr' rU_{jl}^{[s]}(r,r')r' u_{jl}(r') \; .
\end{align}
Here ${\cal D}_{c}$ denotes a second order
differential operator which includes
the point--Coulomb contribution, with the Sommerfeld parameter
$\eta$ given by $\eta=m\beta/ \hbar^2 k$.
Two linearly independent homogeneous solutions to Eq.~\eqref{schr2} 
are the regular (${F}_l$) and irregular (${G}_l$) Coulomb wavefunctions
which satisfy ${\cal D}_{c} F_l(kr) = {\cal D}_{c} G_l(kr) = 0$.
We adopt phase conventions such that their asymptotic behavior are given by
\begin{align}
\label{ric2}
F_l(z) &\,\rule[-6pt]{.1pt}{14pt}_{\;z\to\infty}
\rightarrow      \sin(z - l\pi/2 - \eta\ln 2z + \sigma_l)\;,
\nonumber \\
G_l(z) &\,\rule[-6pt]{.1pt}{14pt}_{\;z\to\infty}
\rightarrow      \cos(z - l\pi/2 - \eta\ln 2z + \sigma_l)\;.
\end{align}

\subsection{Formal solution}
\label{formal}
We now look for a solution for the scattering wavefunctions in the
presence of the Coulomb term. 
Let us first recall the case where the Coulomb interaction is suppressed.
In such a case, if $\hat V$ represents a short--range potential,
the Lippmann--Schwinger integral equation for scattering
waves $|\psi\rangle$ at a given energy $E$ reads
\begin{equation}
  \label{ls0}
  |\psi\rangle =
  |\phi_0\rangle +
  \hat G_0(E+i\eta)\hat V |\psi\rangle\;,
\end{equation}
where $|\phi_0\rangle$ represents free incoming waves and
${\hat G_0(E+i\eta)=(E+i\eta-\hat K)^{-1}}$, corresponding 
to the free propagator.
Here $\hat K$ is the kinetic energy operator, so that ${\hat K|{\bm k}\rangle = (k^2/2m)|{\bm k}}\rangle$.
To obtain the scattering waves in coordinate space it is customary to evaluate the 
free propagator in coordinate representation, i.e. ${\langle {\bm r}|\hat G_0(E+i\eta)|{\bm r}'\rangle}$. 
Following Joachain~\cite{Joachain1975}, after performing partial wave expansions and subsequent 
contour integrations in the complex $k$--plane it is found that
\begin{equation}
  \label{green0}
  \langle {\bm r}|\hat G_0(E+i\eta)|{\bm r}'\rangle
  =
  \frac{2m}{\hbar^2}
  \sum_{l=0}^{\infty}\sum_{m=-l}^{l} 
  \left ( -\frac{i}{k}\right )
   j_l(kr_{<})h_l^{(+)}(kr_{>}) 
  Y_{lm}(\hat{\bm r})
  Y_{lm}^{*}(\hat{\bm r}')
  \;,
\end{equation}
where $h_l^{(+)}=j_l-in_l$. 
Here $j_l$ and $n_l$ denote spherical Bessel and Neumann functions,
respectively.
Additionally, $r_{<}=\min\{r,r'\}$, while $r_{>}=\max\{r,r'\}$. 
If we now include a Coulomb interaction, then Eq.~\eqref{ls0} for the wavefunction can be cast as
\begin{equation}
  \label{lsc}
  |\psi\rangle =
  |\chi_0\rangle +
  \hat G_C(E+i\eta)\hat U^{[s]} |\psi\rangle\;,
\end{equation}
where $|\chi_0\rangle$ correspond to free incoming Coulomb waves, and
\begin{equation}
  \label{greenc}
  \hat G_C(E+i\eta)=\frac{1}{E+i\eta-\hat K-\hat V_c}\;,
\end{equation}
to the free Coulomb propagator.
In this case, $\hat V_C$ corresponds to the point Coulomb interaction and $U^{[s]}$ 
is defined in Eq.\eqref{us}. The difficulty in this case is that there is no known 
procedure to obtain $\langle {\bm r}'|\hat G_C(E+i\eta)|{\bm r}\rangle$, in analogy to the 
one adopted to obtain Eq.~\eqref{green0} for the propagator. Most of the difficulty 
arises from the fact that $\hat K$ does not commute with $\hat V_C$, preventing 
manageable contour integrations in the complex $k$--plane.

To circumvent the above difficulty with Coulomb interactions, we look for a solution 
for outgoing scattering waves $u_{jl}$ in Eq.~\eqref{schr2}, expressed as the superposition 
of homogeneous and particular solutions in the form
\begin{align}
\label{solution}
u_{jl}(r) = &
\frac{1}{k} F_l(kr) + 
\frac{2m}{\hbar^2}
\iint dr'dr'' 
 G_l^{c(+)}(r,r';k) \left [ 
r'U^{[s]}_{jl}(r',r'') r'' \right ] u_{jl}(r'') \; .
\end{align}
For the construction of a particular solution we pursue the following ansatz for the Coulomb 
propagator $G_l^{c(+)}$ in partial wave $l$,
\begin{equation}
\label{propagator}
G_l^{c(+)}(r,r';k) 
 = 
-
  \frac{i}{k}
F_l(kr_{<}){\cal H}_l^{(+)}(kr_{>}) \;,
\end{equation}
where ${\cal H}_l^{(+)}=F_l-iG_l$.

The validity of this ansatz for $G_l^{c(+)}$ calls for a demonstration. To do so, we verify that 
the formal solution expressed by Eq.~\eqref{solution} for $u_{jl}$, satisfies the integro--differential 
equation in Eq.~\eqref{schr2}. Hence, let us examine the action of ${\cal D}_{c}$ on $F_l$ and the integral 
involving the kernel. Since $F_l$ satisfies ${\cal D}_{c} F_l = 0$, then we just need to focus on 
\begin{equation}
\label{igw}
{\cal Z}(r)\equiv{\cal D}_{c} 
\int_{0}^{\infty}
dr' G_l^{c(+)}(r,r';k) W_{jl}(r') \; ,
\end{equation}
where $W_{jl}(r')$ represents the integral over $r''$ given by
\begin{equation}
W_{jl}(r') \equiv 
\frac{2m}{\hbar^2}
\int_{0}^{\infty}
r'U^{[s]}_{jl}(r',r'')r'' u_{jl}(r'') dr''\;.
\end{equation}
Making explicit $G^{c(+)}_{l}$ defined in Eq.~\eqref{propagator} by splitting the integral over 
$r'$ in Eq.~\eqref{igw} into two sub-intervals, $[0,r]$ and $[r,\infty)$, we get
\begin{eqnarray}
\label{split}
  \int_0^{\infty} G_l^{c(+)}(r,r';k) W_{jl}(r') dr'dr''  &=&  
  -\frac{i}{k} \left [
{\cal H}_l^{(+)}(kr)
  \int_{0}^{r} dr' F_l(kr') W_{jl}(r') \right.
 \nonumber \\
  &+& \left. F_l(kr) \int_{r}^{\infty} dr' {\cal H}_l^{(+)}(kr') W_{jl}(r') 
	\right ]
\end{eqnarray}
Taking derivatives with respect to $r$ and using the Wronskian identity 
\begin{equation}
  \label{wronsk}
F_l(z) {\cal H}_l^{(+)\prime}(z)-F_l'{(z)\cal H}_l^{(+)}(z)= i\;,
\end{equation}
we obtain 
\begin{eqnarray}   
- \frac{\partial^2}{\partial r^2} 
\int_{0}^{\infty} G_l^{c(+)}(r,r';k)  W_{jl}(r') dr' &=&  W_{jl}(r)  
\nonumber \\
&+&\frac{i}{k} 
	\left [  
\frac{\partial^2 {\cal H}_l^{(+)}(kr)}{\partial r^2}
\right . 
\int_{0}^{r} dr' F_l(kr')W_{jl}(r')  
\nonumber \\
&+&\left .\frac{\partial^2 F_l(kr)}{\partial r^2}
\int_{r}^{\infty} dr' {\cal H}_l^{(+)}(kr')  W_{jl}(r') 
	\right ] \nonumber \\
\end{eqnarray}
Combining this result with Eq.~\eqref{split} and considering that
${\cal D}_{c}F_l= {\cal D}_{c}{\cal H}_l^{(\pm)}=0$, we get
\begin{equation}
  {\cal Z}(r) =
{\cal D}_{c} \int G_l^{c(+)}(r,r';k) W_{jl}(r') dr' = W_{jl}(r)\,,
\end{equation}
proving that $u_{jl}$ as given by Eq.~\eqref{solution} constitutes the solution to the wave equation 
\eqref{schr2} for outgoing scattering waves.
\\

An appealing feature of the propagator expressed 
by Eq.~\eqref{propagator} is that it is non--singular,
being a continuous function of $r$ and $r'$. 
The gradient of $G_l^{c(+)}$ is discontinuous 
at the diagonal $r=r'$, although this feature poses no particular drawback. 
Note that Eq.~\eqref{solution} takes the form of a
Lippmann--Schwinger integral equation 
for scattering waves in the presence of Coulomb interaction, 
which we recast as
\begin{equation}
\label{ekernel}
  \int dr'' 
  \left [
  \delta(r-r'') -
  K_{jl}(r,r'')
  \right ] 
  u_{jl}(r'') = 
  \textstyle{\frac{1}{k}} F_l(kr)\;,
\end{equation}
where the kernel $K_{jl}$ is given by
\begin{equation}
\label{kernel}
K_{jl}(r,r'') =
\frac{2m}{\hbar^2}
\int dr' 
 G_l^{c(+)}(r,r';k) 
  \left [ 
r'U^{[s]}_{jl}(r',r'') r'' 
  \right ] \; .
\end{equation}
This kernel contains the nonlocal hadronic interaction 
superposed to the Coulomb--screened electrostatic interaction. 
Note that Eq.~\eqref{ekernel} enables to obtain the actual 
scattering wavefunction, solution of Schr\"odinger's 
integro--differential equation, by means of direct matrix inversion. 
In this context, the solutions for the scattering waves are exact.
The novel feature here is that Coulomb interaction is also
treated exactly.

The solution for $u_{jl}$ from Eq. \eqref{ekernel} enables the 
calculation of the scattering amplitude, which follows from
the asymptotic form of Eq.~\eqref{solution}, where $r$ is taken 
far away from the scattering center. In this limit we have
\begin{equation}
\label{faraway2}
G_l^{c(+)}(r,r';k) \,\rule[-6pt]{.1pt}{14pt}_{\,r\gg r'} \longrightarrow
 -
\frac{i}{k} 
F_l(kr'){\cal H}_l^{(+)}(kr) \; ,
\end{equation}
which once replaced in Eq.~\eqref{solution} for $u_{jl}$ yields
\begin{equation}
\label{match}
k\,u_{jl}(r)
\,\rule[-6pt]{.1pt}{14pt}_{\;r\to\infty}
\rightarrow 
F_l(kr) + \Delta_{jl}  
\left [F_l(kr) \mp iG_l(kr)\right ]
\;,
\end{equation}
with  
\begin{equation}
\Delta_{jl} = - 
\frac{2mi}{\hbar^2} 
\iint r'dr'\,r'' dr''
 F_l(kr') U^{[s]}_{jl}(r',r'')  u_{jl}(r'') \; .
\end{equation}
These last two relations allow independent ways to obtain $\Delta_{jl}$. 
The latter involves direct integration of the wavefunction whereas
the former evaluates asymptotically the ratio
\begin{equation}
\label{asymp}
\Delta_{jl} = \frac{ku_{jl}(r) - F_l(kr)}{F_l(kr) - iG_l(kr) }\;,
\end{equation}
for sufficiently large $r$.
These equivalent forms to calculate $\Delta_{jl}$ serve as 
a means to crosscheck consistency of the solutions.
Once $\Delta_{jl}$ is obtained, the scattering amplitude $f_{jl}$ 
and short--range phase shift $\hat\delta_{jl}$ follow from
\begin{equation}
\Delta_{jl} =
 ikf_{jl} =
\textstyle{\frac12}\left (e^{ 2i\hat\delta_{jl}} - 1 \right )
\;.
\end{equation}

\subsection{Numerical application}
To illustrate the consistency of the solution expressed by Eq.
\eqref{solution} under nonlocal interactions, 
we present applications for $p+^{40}$Ca elastic scattering at 
30.3 and 300 MeV beam energies.
For these examples we choose microscopic optical model 
potentials taken from momentum--space
{\it in-medium} folding calculations, where the mixed
density of the target is folded to the full off--shell $g$ matrix, 
accounting for the Fermi motion of target nucleons \cite{Aguayo2008}. 
The bare nucleon--nucleon interaction used to calculate 
fully off--shell $g$ matrices
is Argonne $v_{18}$ \cite{Wiringa1995}.
The optical potential is then transformed to coordinate 
space as described in Ref.~\cite{Arellano2018},
resulting in nonlocal potentials with intricate structure,
depending on the momentum cutoff used in the Fourier transform.
The Coulomb interaction corresponds to that due to a uniform charge 
distribution.
No localization of hadronic contributions 
is performed at any stage of the calculations.

The numerical implementation of Eq. \eqref{ekernel} follows from
the discretization of $r$ and $r'$ over an $N$--point
uniform mesh, where $r\to r_n=n\,h$, with $h$ a suitable spacing. 
Trapezoidal rule is adequate in this case.
The kernel, function of $r$ and $r'$, becomes a finite $N\times N$ 
matrix which we denote as $\mathbb{K}$. The solution to Eq.~\eqref{ekernel} 
takes the form
\begin{equation}
\label{matrixeq}
\textrm{\bf u} =  (1 - \mathbb{K})^{-1}\textrm{\bf u}_0\;,
\end{equation}
with $\textrm{\bf u}_0$ the unperturbed wave $F_l(kr)/k$, 
while $\textrm{\bf u}$ denotes the scattering wave over 
the discrete mesh. 
In this case we use $N~=~150$, with spacing $h~=~0.1$~fm. 
Note that the scattering wavefunction is fully determined 
from Eq.~\eqref{matrixeq}, 
requiring no normalization to match asymptotic waves. 
Results from this approach (referred in the following 
as Exact Scattering Waves, ESW) are compared with those 
obtained from DWBA98 code \cite{Raynal1998}, 
which provides exact numerical solutions for Schr\"odinger's 
integro--differential nonlocal wave equation. 

\begin{figure}
\resizebox{0.95\linewidth}{!}{%
\includegraphics[angle=-90,origin=c,clip=true]{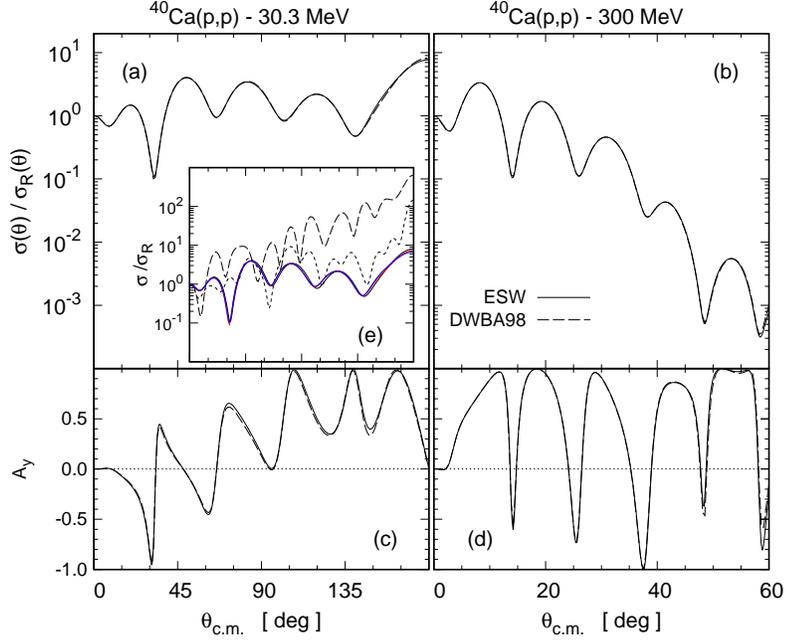}
}
\caption{
  Calculated ratio--to--Rutherford elastic cross section
  (a,b) and analyzing power (c,d) 
  as functions of the center--of--mass scattering angle. 
  Results obtained from microscopic nonlocal 
  potential for $^{40}$Ca$(p,p)$ elastic scattering 
  at 30.3 and 300~MeV. 
  Solid and dashed curves denote results from ESW (this work) 
  and DWBA98, respectively. 
  Inset (e) shows results for $\sigma/\sigma_R$ for
  Perey-Buck--type potential obtained from 
  ESW (solid blue curves), 
  DWBA98 (solid red curve) and
  NLAT (solid black curves).
  Inset (e) also shows results for microscopic optical model
  obtained with NLAT using KD \cite{koning2003} 
  and CH89 \cite{Varner1991} potentials as starting 
  solutions in the iterative procedure, denoted with 
  dashed and dotted curves, respectively.
}
\label{xay}       
\end{figure}

In Fig.~\ref{xay} we show results for the 
ratio--to--Rutherford of the elastic cross section
$\sigma(\theta)/\sigma_R(\theta)$ (a,b) 
and analyzing power $A_y$ (c,d) as functions of 
the scattering angle in the center--of--mass reference frame. 
Frames on the left-hand side correspond to 30.3~MeV proton 
scattering off $^{40}$Ca, 
and those on the right-hand side correspond to 300~MeV. 
Solid curves represent results based on the present approach (ESW), 
while dashed curves represent solutions 
using DWBA98 code \cite{Raynal1998}. 
In the case of 30.3~MeV we observe that 
differences in $\sigma/\sigma_R$ become slightly 
noticed for $\theta_{c.m.}>140^\circ$. 
In the case of the analyzing power, differences are 
quite moderate but enough to distinguish the two approaches. 
Results for proton scattering at 300~MeV are plotted 
up to $\theta_{c.m.}=60^\circ$, 
corresponding to a relatively high momentum transfer of 4~fm$^{-1}$. 
In this case we note that the curves for both 
$\sigma(\theta)/\sigma_R(\theta)$ and $A_y$ overlap almost completely,
illustrating the level of agreement for the two exact approaches.

In the context of the numerical application at 30.3~MeV, we have also 
investigated the use of NLAT code \cite{Titus2016}, developed to solve 
the nonlocal Schr\"odinger equation using an iterative procedure. 
Results from this code using Perey--Buck--type potential in the 
version developed by Tian \textit{et al.} \cite{Tian2015} are in 
reasonable agreement with the ones obtained with ESW and DWBA98 approaches.
This is illustrated in inset (e) of Fig.~\ref{xay}, 
where we plot the ratio--to--Rutherford of the elastic cross section.
Black, red and blue curves denote results for NLAT, DWBA98 and ESW, 
respectively, displaying reasonable agreement among them.
However, when NLAT code is used for the microscopic model 
it fails to solve the nonlocal equation. 
Inset (e) shows results from two trial local potentials 
proposed in the regular input of NLAT. 
One is Koning--Delaroche (KD) potential 
\cite{koning2003} (dashed curve) 
and the other Chapel--Hill potential (CH89) 
\cite{Varner1991} (dotted curve).
As observed, these trial solutions lead to different solutions for
the cross sections, demonstrating the sensitivity to the 
kernel--shape of NLAT approach in its present version. 
It is worth noting that the cross section 
obtained from ESW using the microscopic potential 
is very similar to that from Perey--Buck--Tian nonlocal potential, 
so one would expect KD to be a reasonable trial potential in the 
microscopic case as well. 

\section{Discussion and concluding remarks}
\label{conclu}
The solution embodied by Eq.~\eqref{solution} for the scattering waves in the presence 
of Coulomb interactions is a piece of knowledge overlooked in the field. As demonstrated, 
this equation leads univocally to {\it the solution} for the scattering waves. By contrast, 
any iterative method can always be re-expressed as an infinite series, being also equivalent 
to a perturbative approach. Assessing beforehand its convergence is an issue with no formal 
solution. In order to anticipate the convergence of any iterative method one needs information 
on the initial guess in addition to the structure of the kernel. At the end, their effectiveness 
relies on empirical know--how under controlled scenarios.

  In summary, we have presented an exact solution for the scattering waves off nonlocal optical 
potentials in the presence of long--range Coulomb interaction. The structure of the solution poses 
no restrictions on the type of nonlocality, beam energy nor charge of colliding particles.
Its numerical implementation leads to non--singular finite matrices over a spatial mesh, allowing 
to obtain the scattering waves by direct matrix inversion. When compared to exact solutions of the 
integro--differential Schr\"odinger's equation provided by the DWBA98 code, excellent agreement is 
observed in the calculated scattering observables at nucleon energies of up to 300 MeV. With these 
features, the solution we present provides benchmark solutions to compare with. Since the approach 
we present leads to actual solutions for the scattering waves, it is well suited for distorted--wave 
Born approximation for nuclear reactions. Additionally, the approach presented here is well suited 
for coupled--channels \cite{Arellano2007b}, with extension to inelastic processes 
underway \cite{Nasri2017}.

\section*{Acknowledgement}
H.F.A. thanks colleagues at CEA, Bruy\`eres--le--Ch\^atel, France, 
for their hospitality during his stay where part of this work was done.

\end{document}